\documentclass[12pt,a4paper]{article}
\usepackage{graphicx}
\usepackage[cp1251]{inputenc}
\usepackage{amssymb, amsfonts, amsmath}
\numberwithin{equation}{section}
\date{}
\usepackage{amscd}
\begin{document}

\title{Hamilton-Jacobi and Klein-Gordon-Fock equations for a charged test particle in space-time with simply transitive four-parameter groups of motions}

 \author{V.V. Obukhov}

 \maketitle
 

            vectors and tensors, motion integrals, groups of motions of a space.

            \quad

Tomsk State Pedagogical University (TSPU),  Institute, of Scietific Research and

Development, 60 Kievskaya St., Tomsk, 634041, Russia;

 e.mail: obukhov@tspu.edu.ru.

 \quad

 Tomsk State University of Control Systems and Radio Electronics (TUSUR),

 Laboratory for Theoretical Cosmology, International Center of Gravity and Cosmos,

 36, Lenin Avenue, Tomsk, 634050, Russia

\section{Introduction}

Space-time manifolds, the symmetry of which is due to the presence of vector and tensor Killing fields, are of particular interest in geometry and in the theory of gravity, since the Killing fields define the motion integrals of the geodesic equations.
When the sets of this fields satisfy some additional conditions, solutions of the geodesic equations can be obtained by commutative integration (see, for example, \cite{1}-\cite{7} and bibliography in papers \cite{8}-\cite{9}), or by non-commutative integration (see \cite{11}-\cite{12}). In this case the sets of Killing fields are called complete sets.

The method of  commutative integration (the method of complete separation of variables)  is possible to use,  when the space-time $V_4$ admits a commuting complete set of Killing fields. Noncommutative integration method can be used, for example, when three- or four-parametric groups of motions act simply transitively on the space $V_4$ (or on the hypersurfaces $V_3$ of the space $V_4$).

The set of operators of these groups will be called a non-commuting complete set. Note that presence in the space complete sets of Killing fields is a necessary but not sufficient condition for the existence of the complete sets of integrals of motion of the Hamilton-Jacobi equation for a charged classical test particle, as well as for quantum mechanical equations of motion (the Klein-Gordon-Fock, Dirac-Fock, etc.). Finding metrics and electromagnetic potentials that satisfy sufficient conditions for the existence of motion integrals is a complex of serious classification problems. The solution to these problems should be the following:

1. Finding all non-equivalent sets of space-time metrics, in which
the free Klein-Gordon-Fock, Dirac-Fock and other quantum-mechanical equations admit motion integrals that are linear or squared in momentums.

2. Finding all non-equivalent sets of space-time metrics and potentials of admissible electromagnetic fields, in which the Hamilton-Jacobi Klein-Gordon-Fock, Dirac-Fock equations for a charged test particle admit motion integrals that are linear or squared in momentums.

A.Z. Petrov in the book \cite{13} has obtained all space-time metrics that admit non-commuting complete sets. In the paper \cite {13a} it has been shown that in these spaces the free Klein-Gordon-Fock equation admits non-commuting complete sets of integrals of motion, and the method of noncommutative integration can be applied to this equation. Significant results have also been obtained for the free Dirac-Fock equation (see, for example, \cite{12}).

For stackel spaces (in which the method of commutative integration is applicable) the classification problem for the free Klein-Gordon-Fock equation has until recently considered only on the class of exact solutions of the gravitational field equations. As for the Dirac-Fock equations, for it this classification problem was considered without involving gravitational field equations (see\cite{14}-\cite{15}).

The second group of classification problems in stackel spaces for the Hamilton-Jacobi and Klein-Gordon-Fock equations were also considered until recently almost exclusively on the exact solutions of the gravitational field equations (see, for example, \cite{16}, and the bibliography in the papers \cite {8}-\cite {9}). In the papers \cite{8}-\cite{9} this classification problem has been solved without involving the gravitational field equations. All non-equivalent sets were found, consisting of space-time metrics and potentials of external admissible electromagnetic fields, in which the Hamilton-Jacobi equation admits a complete separation of variables.

The problem of a complete classification of metrics and potentials of the admissible electromagnetic fields, in which the Hamilton-Jacobi and Klein-Gordon-Fock equations admit non-commuting complete sets, is close to the final solution. In papers \cite{17} - \cite{19} all admissible electromagnetic fields are found for the case, when the groups of motions $G_3$ act simply transitively on hypersurfaces  of space-time $V_4$. In paper \cite{19} the case has been considered, when the groups $G_4$ act on $V_3$. It remains to consider the case, when the groups $G_4$ act simply transitively on the space $V_4$.  Thus, the present paper continues the article  \cite{19}, and is its conclusion. The classification of admissible electromagnetic fields on manifolds
$V_4$ with simply transitive groups of motions was given in an algebraic free-coordinate form in
\cite{19a}, \cite{19b}.

We note that the problem of studying spaces with complete sets of Killing fields is still relevant in the theory of gravity and in cosmology.

1.
Spaces with commuting complete sets of Killing fields are of particular interest. The most  important in terms of physics exact solutions of the Einstein equations (the solutions of Schwarzschild \cite{20}, Kerr \cite{21}, Newman-Unti-Tamburino\cite{22}, Friedmann \cite{23} and many others) belong to the class of stackel spaces. In these spaces it is possible to construct exact or reliable approximate solutions of the classical and quantum motion equations of test particles. It makes possible to study in more detail the appropriate solutions of the gravitational field equations.These solutions are also used to build the realistic models of physical processes in strong gravitational fields, and to study cosmological problems. Therefore, stackel spaces attract a fairly wide attention of researchers till now (see, for example, \cite{24}-\cite{36}).

2.
As to spaces with non-commuting complete sets of Killing fields. Homogeneous and plane-wave spaces belong to the class of these spaces. Homogeneous spaces are used for constructing of non-isotropic models of the Universe at the early stages of development. Plane-wave spaces are also of particular interest. They are modeled by spaces, on whose null hypersurfaces $G_3$ groups with a null operator act. A separate class of problems is the study of the intersection of stackel and homogeneous spaces. A large number of articles devoted to the study of spaces with non-commuting complete sets of Killing fields have been published in recent years (see, for example, papers \cite{37}-\cite{43a}).

\section{Algebra of motion integrals}

Let us consider an $n$ - dimensional Riemannian manifold with metric {g} of arbitrary signature, on which the group $G_r$ acts. We denote such a space \quad $V_n(G_r)$.\quad The coordinate indices of variables of the local coordinate system of the space\quad $V_n( G_r)$\quad will be denoted by small Latin letters:\quad $i, j, k, l = 1 \dots n; \quad a, b =1 \dots \dim G_r $.\quad Group indices will be denoted by small Greek letters:\quad $\alpha,\beta,\gamma=1 , \dots \dim G_r.$\quad The summation is performed over the repeated upper and lower indices within the index range.

In the section the subject of our consideration is the set of admissible external electromagnetic fields, in which the classical and quantum equations of motion for a  charged scalar test particle (the Hamilton-Jacobi and Klein-Gordon-Fock equations) admit algebras of motion integrals. These algebras is isomorphic to the algebras of operators of the groups $ G_r$ acting simply transitively on the subspace \quad $V_r$\quad of the space \quad $V_n(G_r)$.

\subsection{Conditions for the existence of the admissible electro-magnetic fields}

Let's start with the Hamilton-Jacobi equation for a charged test particle moving in space \quad $V_n$\quad in the presence of an external electromagnetic field with vector potential $A_i$:
\begin{equation}\label{1}
H = g^{ij}P_iP_j=\lambda = const, \quad P_i=p_i+A_i,\quad p_i=\partial_i\varphi.
\end{equation}
On the space \quad $V_n(G_r)$\quad the group of motions \quad $ G_r (r\leq n)$ \quad acts, defined by the Killing vector fields $\xi^i_\alpha$ . Therefore, the free Hamilton-Jacobi equation admits motion integrals of the form:
\begin{equation}\label{2}
Y_\alpha=\xi_\alpha^i p_i.
\end{equation}
If an equation \eqref{1} with a nonzero electromagnetic field admits $r$ independent first-order integrals of motion, these integrals have the form
\begin{equation}\label{3}
X_\alpha=\xi_\alpha^iP_i +\chi_\alpha
\end{equation}
 (see \cite{13a}, \cite{17}, \cite{19}). The corresponding electromagnetic fields are called admissible.

In contrast to the free Hamilton-Jacobi equation, the equation \eqref{1} generally does not have motion integrals even in the space \quad $V_n(G_r)$.\quad We have to find the sets of equations defines electromagnetic fields, in which the equation \eqref{1} admits an algebra of motion integrals. Obviously, this algebra is isomorphic to the operator algebra of the group \quad $G_r$. \quad
For an arbitrary $r$ the compatibility conditions for these sets of equations were studied in \cite{19}.
If functions $A_i$ are  potentials of the admissible electromagnetic fields, the Klein-Gordon Fock equation:
$$
\hat{H}=g^{ij}\hat{P}_i\hat{P}_j\varphi=m^2\varphi \quad (\hat{P}_i=\hat{p}_i+A_j, \quad \hat{p}_i=-i\hat{\nabla})
$$
admits an algebra of motion integrals of the form:
$$
\hat{X}_\alpha=\xi_\alpha^i\hat{P}_i +\chi_\alpha.
$$
Here \quad $\hat{\nabla}_j$\quad are covariant derivative operators corresponding to partial derivative operators \quad$\hat{\partial}_i=\partial/\partial u_i. $\quad
As shown in the papers \cite{13a}, Klein-Gordon-Fock and Hamilton-Jacobi equations admit algebras of motion integrals in the same admissible electromagnetic fields. Therefore, in the following all calculations are carried out for the Hamilton-Jacobi equation.

Let us prove that an admissible electromagnetic field does not deform the algebra of integrals of motion linear in momenta. The integrals of motion for a free test particle and for a charged one have the forms \eqref{2}, \eqref{3}.
Transform the functions \quad $\chi_\alpha$ \quad as follows:
$$
\chi_\alpha = \omega_\alpha -\xi_\alpha^iA_i.
$$
Then operators $X_\alpha$ have the form:
 \begin{equation}\label{4}
 X_\alpha=\xi_\alpha^ip_i + \omega_\alpha.
 \end{equation}
Let us show that from the conditions
\begin{equation}\label{5}
[X_\alpha,X_\beta]=C^\gamma_{\alpha\beta}X_\gamma
\end{equation}
it follows:
\begin{equation}\label{6}
\omega_\alpha = 0 \Rightarrow \gamma_{\alpha}=-\xi_\alpha^i A_i.
\end{equation}
Substitute \eqref{4} into \eqref{5}. As a result, we get:
$$
[X_\alpha,X_\beta]=C^\gamma_{\alpha\beta}\xi_\gamma^ip_i + \omega_{\beta|\alpha}-\omega_{\alpha|\beta} = C^\gamma_{\alpha\beta}(\xi_\gamma^ip_i+\omega_\gamma) \quad
$$
$$
(|_\alpha =\xi_\alpha^i\partial_i. \quad \omega_\alpha = \xi_\alpha^i\omega_i  = \xi_\alpha^a \omega_a).
$$
Thus, the functions $ \omega_\alpha $ obey the equations:
$$
\omega_{\beta|\alpha}-\omega_{\alpha|\beta} = C^\gamma_{\alpha\beta}\omega_\gamma\quad \rightarrow \quad \omega_{a,b}=\omega_{b,a}\quad \rightarrow \quad \omega_a=\omega_{,a}.
$$
By by gauge transformations of the potential, the function \quad $\omega$\quad can be turned to zero:\quad$ \omega = 0 $.\quad

  We find the conditions that admissible electromagnetic fields must satisfy.
The equation \eqref{1} admits an integral of motion of the form \eqref{2} if $H$ and $Y_\alpha$ commute with respect to Poisson brackets:
\begin{equation}\label{7}
\frac{\partial H}{\partial p_i}\frac{\partial Y_\alpha}{\partial x^i} - \frac{\partial H}{\partial x^i}\frac{\partial Y_\alpha}{\partial p_i}= (g^{il}{\xi^{j}_\alpha}_{,l}+g^{jl}{\xi^{i}_\alpha}_{,l}-g^{ij}_{,l}\xi_\alpha^l)P_i P_j + 2g^{i\sigma}(\xi^{j}_\alpha F_{ji}+(\xi_{\alpha}^\beta A_\beta)_{,i})P_\sigma=0,
\end{equation}
$$F_{ij}=A_{j,i}-A_{i,j}.$$
The \eqref{3} equation must be an identity with respect to momenta. That's why
the functions $\xi^{j}_\alpha$ satisfy the equations:
$$
g^{il}{\xi^{j}_\alpha}_{,l}+g^{jl}{\xi^{i}_\alpha}_{,l}-g^{ij}_{,l}\xi_\alpha^l=0,
$$
and they are the Killing vectors. Equations \eqref{3} also imply:
\begin{equation}\label{8}
(\xi_{\alpha}^j A_j)_{,i} = \xi^{j}_\alpha F_{ij}.
\end{equation}
The sets of defining equations for admissible electromagnetic fields is obtained.

\subsection{Defining equations}

Let us find the conditions that the components of the vectors of the admissible electromagnetic field must satisfy if the group $G_4$ acts on the space-time \quad $V_4(G_4)$\quad simply transitively \quad $(det||\xi_\alpha^i|| \ne 0).$ \quad
In this case the Killing vector can be chosen as the vector of a nonholonomic nonorthogonal tetrad. Denote the tetrad dual to it by $\xi^\alpha_i$. The matrix composed of the components\quad  $\xi^\alpha_i$\quad is inverse to the matrix \quad $||\xi_\alpha^i||$:
$$
\xi_\alpha ^i \xi^\beta _i = \delta_\alpha^\beta.
$$
The tetrad components of the metric tensor, and the tetrad components of the electromagnetic field vector potential have the form \cite{19}:
\begin{equation}\label{9}
\mathbf{G}_{\alpha\beta} = \xi_\alpha^i \xi_\beta^j g_{ij}, \quad \mathbf{G}^{\alpha\beta} = \xi^\alpha_i \xi^\beta_j g^{ij}, \quad \mathbf{A}_\alpha = \xi_\alpha ^i A_i, \quad \mathbf{A}^\alpha = \xi^\alpha_i A^i.
\end{equation}
Using this tetrad, the Killing equations can be represented as:
\begin{equation}\label{10}
\mathbf{G}^{\alpha\beta}_{|\gamma}=\mathbf{G}^{\alpha\tau}C^\beta_{\tau\gamma} +\mathbf{ G}^{\beta\tau}C^\alpha_{\tau\gamma},
\quad \mathbf{G}_{\alpha\beta|\gamma}=\mathbf{G}_{\alpha \tau}C^\tau_{\gamma\beta} +\mathbf{G}_{\beta \tau}C^\tau_{\gamma\alpha},
\end{equation}
where denoted: \quad $|_\gamma = \xi_\gamma^i\partial_i $.
Let us show (see \cite{18}, \cite{19}) that in this case the defining equations \eqref{8} can be written as follows:
\begin{equation}\label{11}
  \mathbf{A}_{\alpha|\beta} = C_{\beta\alpha}^\gamma\mathbf{A}_\gamma.
\end{equation}
$C_{\alpha\beta}^\gamma$ \quad are the structural constants of the group \quad $G_4$,\quad defining the commutation relations of the group operators:
$$
 [\hat{Y}_\alpha,\hat{Y}_\beta]=C_{\beta\alpha}^\gamma \hat{Y}_\gamma.
$$
Indeed, the system of equations \eqref{5} can be represented as:
$$
\xi_{\beta}^i(\xi_{\alpha}^j A_j)_{,i} = \xi_{\beta}^i\xi^{j}_\alpha F_{ij}.
$$
We transform the right side of the equality as follows:
$$
\xi_{\beta}^i\xi^{j}_\alpha F_{ij} = \mathbf{A}_{\alpha|\beta} -\mathbf{A}_{\beta|\alpha} + C_{\alpha\beta}^\gamma\mathbf{A}_\gamma,
$$
whence follows \eqref{11}.
The set of equations \eqref{11} is compatible due to the Bianchi identities.
Note that the conditions \eqref{11} are also valid in the case when the group $G_r$ acts simply transitively on the non null subspace $V_r$ of the space $V_n$ (see \cite{19}).

As an isometry group $G_4$ acts on $V_4$ simply transitively, then there is a diffeomorphism
between $G_4$ and $V_4$. Thus, the metric on $V_4$ is a left-invariant metric on the Lie group
$G_4$, and the Killing vectors $\xi_{\alpha}^i$ can be associated with the right-invariant vector
fields. On the other hand, the admissibility condition \eqref{12} is equivalent to the requirement
that the Lie derivative of the potential form $A = A_i dx^i$ with respect vector fields $\xi_{\alpha}^i$ vanishes.

So, the classification of admissible electromagnetic fields on manifolds
$V_4$ with simply transitive groups of motions is in fact reduced to the classification of left-invariant 1-forms
on four-dimensional Lie groups. This problem has been solved in \cite{19a}. The general form of such a left-invariant 1-form $A$ on $G_4$ is
\begin{equation}\label{12}
A_i = \alpha_\beta \sigma_i^\beta \quad (\alpha_\beta =const)
\end{equation}
where  $ \sigma_\alpha = \sigma_i^\beta dx^i$ are left-invariant 1-forms on $G_4$ ($ \sigma_\alpha ^i$ is the
tetrad co-vectors $e_\alpha^i$).

Note, that this classification was given in an algebraic free-coordinate form, while the list of admissible electromagnetic fields in the present paper is maid in coordinates used in Petrov's book \cite{13}. Due to \eqref{12} it is possible to find the metric tensor of the space $V_4$ with the group $G_4$   without solving the Killing equations \eqref{9}, using the solutions of the defining system.
Indeed, as proved in \cite{19a}, the holonomous (metric) components of the admissible electromagnetic potential can be represented in the form:
\begin{equation}\label{13}
A_i = \alpha_\beta e^\alpha_i\quad (g_{ij}= \eta_{\alpha\beta}e^\alpha_ie^\beta_j, \quad \eta_{\alpha\beta}= const).  \end{equation}
On the other hand, we can express $A_i$ through the solution of the set \eqref{11} as follows:
$$
A_i = \xi_i^\alpha\mathbf{A}_{\alpha}\quad \Rightarrow \quad \xi_i^\alpha\mathbf{A}_{\alpha} - \alpha_\beta e^\alpha_i = 0.
$$
By setting the coefficients in front of the essential parameters $\alpha_\alpha$ to zero, we find the vectors of the tetrad \quad $e^\alpha_i,\quad $  and recover the metric tensor $g_{ij}$.

The classification of metrics and electromagnetic fields that are invariant under the action of the group $G_r$ must be carried out in three stages.

   1. The canonical coordinate system is selected at the first stage. This  system linked to the subspace $V_r$, in which the solutions of the equations of the structure\eqref{13} have the simplest form;

   2. The Killing equations \eqref{10} are integrated at the second stage;

   3. The defining equations \eqref{11} are integrated at the third stage.

For pseudo-Riemannian space $V_4$ with the simply transitive action of the translation group $G_4$ the problems of the first two stages were solved by A.Z. Petrov (\cite{13}). In our papers (\cite{17}-\cite{19}) we have solved the problems of the third stage for the case, when the group $G_r$ acts simply transitively on hyperspaces of the space-time. The case, when the groups $G_4$ act simply transitively on the whole space-time, was studied in \cite{13a} and complete classification in an algebraic free-coordinate form has been obtained. The present article is devoted to the final solution of this classification problem in coordinate form.

\section{Admissible electromagnetic fields}

A.Z. Petrov in the book \cite{13} suggested the following classification of the spaces $V_4(G_4)$ with simply transitive groups of motions $G_4$.

1. Spaces $V_4(G_4)$ with solvable groups $G_4$ containing non-Abelian subgroups $G_3$ (groups \quad $G_4(I) (c \ne 1),\quad G_4(I) (c=1), \quad G_4(II - V)).$

2. Spaces $V_4(G_4)$ with solvable groups $G_4$ containing Abelian subgroups $G_3$ (groups \quad $G_4(VI)_1 - G_4(VI)_4 $).

3. Spaces $V_4(G_4)$ with non-solvable groups \quad (groups \quad$G_4(VII),\quad G_4(VIII)).$

When finding nonequivalent solutions of the constitutive equations \eqref{12}, we will follow this classification, using the notation introduced by Petrov and the group operators found by him. We also present the components of the non-orthogonal tetrad \quad $e^i_\alpha, \quad e_i^\alpha$.

\subsection{Solvable $G_4$ not containing an Abelian subgroup $G_3$}

To find admissible electromagnetic fields, it is necessary to integrate the set of defining equations \eqref{11}. The procedure for integrating these equations is the same for all $G_4$ groups acting simply transitively on the space-time. Therefore, we present the solution scheme only for the group \quad $G_4(I_1) \quad (c\ne 1) $. For all other groups, we confine ourselves to giving the explicit form of the system \eqref{11} and its solution.

\subsubsection{Group $G_4(I_1) \quad (c\ne 1)$}

We present the components of the Killing vectors \quad $\xi_\alpha^i$ \quad and the components of the dual tetrad \quad $\xi^\alpha_i$ \quad ($\xi_\alpha^i \xi^\beta_i = \delta^ \beta_\alpha )$ \quad in the form:
\begin{equation}\label{14}
||\xi_\alpha^i||=\begin{pmatrix} 0 & 1 & 0 & 0
\\0 & 0 & 1 & 0 \\ -1 & u^3 & 0 & 0 \\ \varepsilon u^1 & c u^2 & u^3 &1 \end{pmatrix},\quad ||\xi^\alpha_i||=\begin{pmatrix}u^3 & 0 & -1 & 0 \\1 & 0 & 0 & 0 \\ 0 & 1 & 0 & 0 \\ -(\varepsilon u^1 u^3 +c u^2) & -u^3 & \varepsilon u^1 & 1  \end{pmatrix},
\end{equation}
where \quad $c=const, \quad \varepsilon = c-1,$ \quad subscripts number the lines. The following components of the structural constants are nonzero:
$$
C^1_{14} = c, \quad C^1_{23}=1, \quad C^2_{24}=1, \quad C^3_{34}= \varepsilon.
$$
Let us represent the system of constitutive equations \eqref{12} as:
\begin{equation}\label{15}
\left\{\begin{array}{ll}
\mathbf{A}_{1,2} = \mathbf{A}_{1,3}=\mathbf{A}_{1,1} =0\quad
\mathbf{A}_{1,4}= -c \mathbf{A}_{1}; \cr
\mathbf{A}_{2,2} = \mathbf{A}_{2,3}=0,\quad \mathbf{A}_{2,1} = \mathbf{A}_{1}, \quad
\mathbf{A}_{2,4}=-(\varepsilon u^1\mathbf{A}_{1}+\mathbf{A}_{2}); \cr
\mathbf{A}_{3,2} =\mathbf{A}_{3,1} = 0, \quad \mathbf{A}_{3,3}=\mathbf{A}_{1},\quad
+ \mathbf{A}_{3,4}= -(u^3 \mathbf{A}_{1}+\varepsilon\mathbf{A}_{3}); \cr
\mathbf{A}_{4,2} = c\mathbf{A}_{1}, \quad \mathbf{A}_{4,3}=\mathbf{A}_{2},\quad
\mathbf{A}_{4,1}= (cu^3\mathbf{A}_{1}-\varepsilon\mathbf{A}_{3}), \quad \cr
\mathbf{A}_{4,4}+\varepsilon u^1\mathbf{A}_{4,1}+ cu^2 \mathbf{A}_{4,2} +u^3 \mathbf{A}_{4,3} = 0.
\\\end{array}\ \right.\end{equation}
From the equations \eqref{15} it follows:
\begin{equation}\label{16}
\mathbf{A}_{1} = \alpha_2 \exp(-cu^4).
\end{equation}
Using \eqref{16} we get:
$$
\mathbf{A}_{2} = \alpha_2 u^1\exp(-cu^4) + A,
$$
where $A$ is a function of the variable $u^4$. The form of this function is determined from the set of equation \eqref{15}. As a result, we find:
$$
\mathbf{A}_{2} = \alpha_2 u^1\exp(-cu^4) + \alpha_3 \exp(-u^4).
$$
Proceeding in a similar way, we find the component $ \mathbf{A}_{3}$.
$$
\mathbf{A}_{3} = \alpha_2 u^3 \exp(-cu^4) + \alpha_1\exp(-\varepsilon u^4).
$$
Let us substitute these relations into the equations \eqref{15}. As a result, we get the component$\mathbf{A}_{4}$:
\begin{equation}\label{17}
 \mathbf{A}_{4} = \alpha_2 (c u^2 + u^1u^3)\exp(-cu^4)- c\alpha_1 u^1 \exp(-\varepsilon u^4)+ \alpha_3 u^3 \exp(-u^4) +\alpha_4.
\end{equation}
Here and everywhere in the text:
$$
\alpha_i = const, \quad a, b = 1, 2, 3.
$$
From the relations \quad $A_i =\xi_i^\alpha \mathbf{A}_{\alpha} $ \quad we find the holonomic components of the vector potential:
\begin{equation}\label{18}
A_{1} = \alpha_1\exp(-\varepsilon u^4), \quad A_{2}=\alpha_2\exp(-cu^4),
\end{equation}
$$\quad A_{3}= \alpha_2u^1\exp(-cu^4)+ \alpha_3\exp(-u^4),\quad A_{4} = \alpha_4.$$

In conclusion, we present a matrix composed from the components of the $e^\alpha_i, \quad e_\alpha^i$ tetrad vectors in the form:(corrected)
\begin{equation}\label{19}
||e^\alpha_i||=\begin{pmatrix}\exp(-\varepsilon u^4) & 0 & 0 & 0
\\0 &\exp(-c u^4) &0& 0 \\ 0 &  u^1 \exp(-c u^4) &\exp(-u^4) &0 \\ 0 & 0 & 0 & 1, \end{pmatrix},
\end{equation}
$$
||e_\alpha^i||=\begin{pmatrix}\exp(\varepsilon u^4) & 0 & 0 & 0
\\0 &\exp(cu^4) &0& 0 \\ 0 & -u^1 \exp(u^4) &\exp(u^4) &0 \\ 0 & 0 & 0 &1 \end{pmatrix}.
$$
Further, when considering the remaining groups, the details of the calculations are omitted.

\subsubsection{Group $\bf{G_4(I_2) \quad (c=1)}$}

The components of the Killing vectors \quad $\xi_\alpha^i$ \quad and the components of the dual tetrad \quad $\xi^\alpha_i$ \quad as well as non-zero components of structure constants \quad $C^\alpha_{\beta\gamma}$\quad can be represent in the form:
\begin{equation}\label{20}
||\xi_\alpha^i||=\begin{pmatrix} 0 & 1 & 0 & 0
\\0 & 0 & 1 & 0 \\ -1 & u^3 & 0 & 0 \\ 0 & u^2 & u^3 &1 \end{pmatrix}, ||\xi^\alpha_i||=\begin{pmatrix}u^3& 0 & -1 & 0
\\1 &0 &0& 0 \\ 0 & 1 & 0 &0 \\ -u^2 & -u^3 & 0 & 1  \end{pmatrix},
\end{equation}
$$
C^1_{14} = 1, \quad C^1_{23}=1, \quad C^2_{24}=1.
$$
We represent the system of defining equations \eqref{12} as:
\begin{equation}\label{21}
\left\{\begin{array}{ll}
\mathbf{A}_{i,2} = \delta^4_i\mathbf{A}_{1},\quad \cr
\mathbf{A}_{i,1} - u^3 \mathbf{A}_{i,2}=\delta^2_i\mathbf{A}_{1}, \cr
\mathbf{A}_{i,3} =\delta^4_i\mathbf{A}_{2} +\delta^3_i\mathbf{A}_{1}, \cr
c\mathbf{A}_{i,1}+ u^2 \mathbf{A}_{i,2} +u^3 \mathbf{A}_{i,3} + \mathbf{A}_{i,4}+\delta^1_i\mathbf{A}_{1} +\delta^2_i\mathbf{A}_{2}=0.
\\\end{array}\ \right.\end{equation}
From the equations \eqref{21} it follows:
\begin{equation}\label{22}
\mathbf{A}_{1} = \alpha_2 \exp(-u^4), \quad \mathbf{A}_{2} =(\alpha_2 u^1+\alpha_3)\exp(-u^4), \quad
\mathbf{A}_{3} = -\alpha_1 + \alpha_2 u^3 \exp(-u^4),
\end{equation}
$$\mathbf{A}_{4} =( \alpha_2(u^1u^3 + u^2) +\alpha_3 u^3 )\exp(-u^4) +\alpha_4.$$
The holonomic components of the vector potential \quad $A_i =\xi_i^\alpha \mathbf{A}_{\alpha}$,\quad and the matrixes \quad $||e^\alpha_i||, \quad||e_\alpha^i||$ \quad have the form:
\begin{equation}\label{23}
A_{1} = \alpha_1, \quad A_{2}=\alpha_2\exp(-u^4),\quad A_{3}= (\alpha_2u^1 +\alpha_3 )\exp(-u^4),\quad A_{4} =\alpha_4;
\end{equation}
\begin{equation}\label{24}
||e^\alpha_i||=\begin{pmatrix}1 & 0 & 0 & 0
\\0 &\exp(-u^4) & 0 & 0 \\ 0 & u^1\exp(-u^4) &\exp(-u^4) &0 \\ 0 & 0 & 0 & 1 \end{pmatrix},
\end{equation}
$$
||e_\alpha^i||=\begin{pmatrix}1 & 0 & 0 & 0
\\0 &\exp(u^4) & 0 & 0 \\ 0 & -u^1\exp(u^4) &\exp(u^4) &0 \\ 0 & 0 & 0 & 1 \end{pmatrix}.
$$
\subsubsection{Group $G_4(II)$}

The components of the Killing vectors \quad $\xi_\alpha^i$ \quad and the components of the dual tetrad \quad $\xi^\alpha_i$ \quad as well as non-zero components of structure constants \quad $C^\alpha_{\beta\gamma}$\quad can be represent in the form:
\begin{equation}\label{25}
||\xi_\alpha^i||=\begin{pmatrix}0 & 1 & 0 & 0
\\0 & 0 & 1 & 0 \\ -1 & u^3 & 0 & 0 \\u^1 & \frac{1}{2}(4u^2 +{u^1}^2) & (u^3 - u^1) & 1 \end{pmatrix},\quad
\end{equation}
$$
||\xi^\alpha_i||=\begin{pmatrix}u^3& 0 & -1 & 0
\\1 &0 &0& 0 \\ 0 & 1 & 0 &0 \\-(u^1u^3 + 2u^2 +\frac{1}{2}{u^1}^2) & (u^1 - u^3) &  u^1 & 1\end{pmatrix},
$$
$$
C^1_{14} = 2, \quad C^1_{23}=1, \quad C^2_{24}=1, \quad C^2_{34}=1,\quad C^3_{34}=1.
$$
We represent the system of defining equations \eqref{12} as:
\begin{equation}\label{26}
\left\{\begin{array}{ll}
\mathbf{A}_{i,2} = 2\delta^4_i\mathbf{A}_{1},\quad \cr
\mathbf{A}_{i,1} - u^3 \mathbf{A}_{i,2}=\delta^2_i\mathbf{A}_{1} -\delta^4_i(\mathbf{A}_{2} +\mathbf{A}_{3}), \cr
\mathbf{A}_{i,3} =\delta^4_i\mathbf{A}_{2} +\delta^3_i\mathbf{A}_{1}, \cr
u^1\mathbf{A}_{i,1}+ \frac{1}{2}(4u^2 +{u^1}^2)\mathbf{A}_{i,2} +(u^3-u^1) \mathbf{A}_{i,3} + \mathbf{A}_{i,4}\cr =-(2\delta^1_i\mathbf{A}_{1}+\delta^2_i\mathbf{A}_{2} +\delta^3_i(\mathbf{A}_{2} +\mathbf{A}_{3})).
\\\end{array}\ \right.\end{equation}
From the equations \eqref{26} it follows:
\begin{equation}\label{27} \left\{\begin{array}{ll}
\mathbf{A}_{1} = \alpha_1 \exp(-2u^4), \cr
\mathbf{A}_{2} =u^1 \mathbf{A}_{1} + \alpha_2 \exp(-u^4),\cr
\mathbf{A}_{3} = u^3\mathbf{A}_{1}-\alpha_2 u^4 \exp(-u^4)+\alpha_3 \exp(-u^4),\cr
\mathbf{A}_{4} =(u^1u^3 + 2u^2-\frac{1}{2}{u^1}^2)\mathbf{A}_{1}+\alpha_2 (u^3 + u^1u^4 - u^1) \exp(-u^4)\cr-\alpha_3 u^1\exp(-u^4)+\alpha_4. \\\end{array}\ \right.
\end{equation}
The holonomic components of the vector potential \quad $A_i =\xi_i^\alpha \mathbf{A}_{\alpha}$,\quad and the matrixes \quad $||e^\alpha_i||, \quad ||e_\alpha^i||$ \quad have the form:
\begin{equation}\label{28}
A_{1} = (\alpha_2 u^4-\alpha_3) \exp(-u^4), \quad
A_{2}=\alpha_1\exp(-2u^4), \quad
\end{equation}
$$
A_{3}= \alpha_1 u^1 \exp(-2u^4) +\alpha_2\exp(-u^4), \quad
A_{4} = \alpha_4,
$$
\begin{equation}\label{29}
||e^\alpha_i||=\begin{pmatrix} 0 &u^4\exp(-u^4) & -\exp(-u^4) & 0
\\\exp(-2u^4) & 0 & 0 & 0 \\ u^1\exp(-2u^4) &\exp(-u^4)& 0  &0 \\ 0 & 0 & 0 & 1 \end{pmatrix},
\end{equation}
$$
||e_\alpha^i||=\begin{pmatrix} 0 &\exp 2u^4 & 0 & 0
\\ 0 & -u^1\exp u^4 & \exp u^4 & 0
\\ -\exp u^4 &-u^1u^4\exp u^4& u^4\exp u^4  & 0 \\ 0 & 0 & 0 & 1 \end{pmatrix}.
$$

\subsubsection{Group $G_4(III)$}

The components of the Killing vectors \quad $\xi_\alpha^i$ \quad and the components of the dual tetrad \quad $\xi^\alpha_i$  \quad as well as non-zero components of structure constants \quad $C^\alpha_{\beta\gamma}$\quad can be represent in the form\quad ($\alpha=const$):
\begin{equation}\label{30}
||\xi_\alpha^i||=\begin{pmatrix}0 & 1 & 0 & 0
\\0 & 0 & 1 & 0 \\ -1 & u^3 & 0 & 0 \\
(2u^1\cos{a}-u^3) & (2u^2\cos{a} + \frac{1}{2}({u^3}^2 -{u^1}^2)) & u^1 & 1 \end{pmatrix},\quad
\end{equation}
$$
||\xi^\alpha_i||=\begin{pmatrix}u^3& 0 & -1 & 0
\\1 &0 &0& 0 \\ 0 & 1 & 0 &0 \\-2(u^1u^3 +u^2)\cos{\alpha} -\frac{1}{2}({u^1}^2+{u^3}^2) & -u^1 & (2u^1\cos{\alpha} - u^3) & 1 \end{pmatrix},
$$
$$
C^1_{14} = 2\cos{\alpha}, \quad C^1_{23}=1, \quad C^3_{24}=1, \quad C^2_{34}= -1,\quad C^3_{34}=2\cos{\alpha}.
$$
We represent the system of defining equations \eqref{12} as:
\begin{equation}\label{31}
\left\{\begin{array}{ll}
\mathbf{A}_{i,2} = 2\cos{\alpha}\delta^4_i\mathbf{A}_{1}, \cr
\mathbf{A}_{i,1} - u^3 \mathbf{A}_{i,2}=\delta^2_i\mathbf{A}_{1} +\delta^4_i(\mathbf{A}_{2} - 2\mathbf{A}_{3}\cos{\alpha}), \cr
\mathbf{A}_{i,3} =\delta^3_i\mathbf{A}_{1} + \delta^4_i\mathbf{A}_{3}, \cr
(2u^1\cos{\alpha} -u^3)\mathbf{A}_{i,1}+ (2u^2\cos{\alpha} + \frac{1}{2}({u^3}^2 -{u^1}^2))\mathbf{A}_{i,2} +u^1 \mathbf{A}_{i,3} + \mathbf{A}_{i,4}\cr
=-2\cos{\alpha}\delta^1_i\mathbf{A}_{1} -\delta^2_i\mathbf{A}_{3} +\delta^3_i(\mathbf{A}_{2} - 2\cos{\alpha}\mathbf{A}_{3})).
\\\end{array}\ \right.\end{equation}
From the equations \eqref{31} it follows:
\begin{equation}\label{32} \left\{\begin{array}{ll}
\mathbf{A}_{1} = \alpha_1\exp(-2u^4\cos \alpha),\cr
\mathbf{A}_{2} = u^1\mathbf{A}_{1} + (A =\exp(-u^4\cos \alpha)(\alpha_2\sin(u^4\sin\alpha) + \alpha_3\cos(u^4\sin\alpha))),\cr
\mathbf{A}_{3} = u^3\mathbf{A}_{1} + (B = \exp(-u^4\cos \alpha)(\alpha_2\sin(u^4\sin\alpha-\alpha) + \alpha_3\cos(u^4\sin\alpha -\alpha))),\cr
\mathbf{A}_{4} = \mathbf{A}_{1}(2u^2\cos{\alpha} + \frac{1}{2}({u^1}^2+{u^3}^2))) +u^1 A +(2 u^1\cos{\alpha} -u^3)B.
\\\end{array}\ \right.
\end{equation}
Here
$$
A =\exp(-u^4\cos \alpha_1)(\alpha\sin(u^4\sin\alpha +\alpha_2),
$$
$$
B = \exp(-u^4\cos \alpha)(\alpha_1(\sin(u^4\sin\alpha-\alpha +\alpha_2)
$$
Then the holonomic components of the vector potential \quad $A_i =\xi_i^\alpha \mathbf{A}_{\alpha}$,\quad and the matrixes \quad $||e^\alpha_i||, \quad ||e_\alpha^i||$ \quad have the form:
\begin{equation}\label{33}\left\{\begin{array}{ll}
A_{1} = \exp(-u^4\cos \alpha)(\alpha_1\sin(u^4\sin\alpha-\alpha) + \alpha_2\cos(u^4\sin\alpha -\alpha)),\cr
A_{2}=\alpha_3\exp(-2u^4\cos\alpha), \cr
A_{3}=  u^1\alpha_3 \exp(-2u^4\cos\alpha) + \exp(-u^4\cos \alpha)(\alpha_1\sin(u^4\sin\alpha) +\cr \alpha_2\cos(u^4\sin\alpha)), \cr
A_{4} = 0; \\\end{array}\ \right.
\end{equation}
\begin{equation}\label{34}
||e^\alpha_i||=\exp(-u^4\cos \alpha)\hat{E},\quad ||e_\alpha^i||=\exp(u^4\cos \alpha)\hat{E}^{-1}.\end{equation}
$$\quad \hat{E} = \begin{pmatrix}0 & \sin(u^4\sin \alpha - \alpha) & \cos(u^4\sin \alpha - \alpha)  & 0\\
\exp(-u^4\cos \alpha) &0 & 0 & 0 \\
u^1\exp(-u^4\cos \alpha) & \sin(u^4\sin \alpha) & \cos(u^4\sin \alpha) &0 \\
0 & 0 & 0 & \exp(u^4\cos \alpha)\end{pmatrix},$$
$$\quad \hat{E}^{-1} = \begin{pmatrix}
0 & \exp(u^4\cos \alpha) & 0 & 0\\
-\frac{\cos(u^4\sin \alpha)}{\sin \alpha} & -u^1\frac{\cos(u^4\sin \alpha - \alpha)}{\sin \alpha}&\frac{\cos(u^4\sin \alpha - \alpha)}{\sin \alpha}  & 0 \\
\frac{\sin(u^4\sin \alpha)}{\sin \alpha} & u^1\frac{\sin(u^4\sin \alpha - \alpha)}{\sin \alpha}&-\frac{\sin(u^4\sin \alpha - \alpha)}{\sin \alpha}  & 0\\

0 & 0 & 0 & \exp(-u^4\cos \alpha) \end{pmatrix}.$$

\subsubsection{Group $G_4(IV)$}

The components of the Killing vectors \quad $\xi_\alpha^i$ \quad and the components of the dual tetrad \quad $\xi^\alpha_i$  \quad as well as non-zero components of structure constants \quad $C^\alpha_{\beta\gamma}$\quad can be represent in the form:
\begin{equation}\label{35}||\xi_\alpha^i||=\begin{pmatrix}0 & 0 & 1 & 0
\\0 & 1 & 0 & 0 \\ -1 & u^2 & 0 & 0 \\
0 & 0 & u^3 & 1 \end{pmatrix},\quad||\xi^\alpha_i||=\begin{pmatrix}0 & u^2 & -1 & 0
\\0 & 1 & 0 & 0 \\ 1 & 0 & 0 &0 \\-u^3 & 0 & 0 & 1\end{pmatrix},
\end{equation}
$$
C^2_{14} = 1, \quad C^2_{23}=1.
$$
Let's find the system of defining equations \eqref{12}:
\begin{equation}\label{36}
\left\{\begin{array}{ll}
\mathbf{A}_{1,i} = - \delta^4_i\mathbf{A}_{1},\quad \cr
\mathbf{A}_{2,2}= \mathbf{A}_{2,3}=\mathbf{A}_{2,4} =0, \quad \mathbf{A}_{2,1}=\mathbf{A}_{2}, \cr
\mathbf{A}_{3,3} = \mathbf{A}_{3,4} = 0,\quad \mathbf{A}_{3,2}= \mathbf{A}_{2}\quad  \mathbf{A}_{3,1} = u^2\mathbf{A}_{2}, \cr
\mathbf{A}_{4,1}=u^2\mathbf{A}_{4,2}, \quad u^3\mathbf{A}_{4,3} + \mathbf{A}_{4,4}=0, \quad \mathbf{A}_{4,3}=\mathbf{A}_{1}, \quad \mathbf{A}_{4,2}=0.
\\\end{array}\ \right.\end{equation}
We integrate the system of equations \eqref{43}. As a result, we get the following solutions:
\begin{equation}\label{37}
\mathbf{A}_{1} = \alpha_3 \exp(-u^4 ), \quad
\mathbf{A}_{2} =\alpha_2\exp(u^1), \quad
\mathbf{A}_{3} = \alpha_2 u^2 \exp(u^1)-\alpha_3,\quad
\end{equation}
$$
\mathbf{A}_{4} = u^3\mathbf{A}_{1} + \alpha_4.
$$
The holonomic components of the vector potential \quad $A_i =\xi_i^\alpha \mathbf{A}_{\alpha}$,\quad and the matrixes \quad $||e^\alpha_i||, \quad ||e_\alpha^i||$ \quad have the form:
\begin{equation}\label{38}
A_{1} = \alpha_1, \quad A_{2}=\beta\exp(u^1),A_{3}= \alpha \exp(-u^4), \quad A_{4} =0;
\end{equation}
$$
$$
\begin{equation}\label{39}
||e_\alpha^i||=\begin{pmatrix}1 & 0 & 0 & 0
\\0 &\exp(-u^1) & 0 & 0 \\ 0 & 0 &\exp(u^4) &0 \\ 0 & 0 & 0 & 1 \end{pmatrix},
||e^\alpha_i||=\begin{pmatrix}1 & 0 & 0 & 0
\\0 &\exp(u^1) & 0 & 0 \\ 0 & 0 &\exp(-u^4) &0 \\ 0 & 0 & 0 & 1 \end{pmatrix}.
\end{equation}

\subsubsection{Group $G_4(V)$}

The components of the Killing vectors \quad $\xi_\alpha^i$ \quad and the components of the dual tetrad \quad $\xi^\alpha_i$  \quad as well as non-zero components of structure constants \quad $C^\alpha_{\beta\gamma}$\quad can be represent in the form:
\begin{equation}\label{40}
||\xi_\alpha^i||=\begin{pmatrix}0 & 1 & 0 & 0
\\0 & 0 & 1 & 0 \\ -1 & u^2 & u^3 & 0 \\
0 & -u^3 & u^2 & 1 \end{pmatrix},\quad||\xi^\alpha_i||=\begin{pmatrix} u^2 & u^3 & -1 & 0
\\1 & 0 & 0 & 0 \\ 0 & 1 & 0 & 0 \\u^3 & -u^2 & 0 & 1\end{pmatrix},
\end{equation}
$$
C^1_{13} = 1, \quad C^2_{14} = 1, \quad C^2_{23}=1, , \quad C^1_{24}=-1.
$$
Let's find the system of defining equations \eqref{12}:
\begin{equation}\label{41}
\left\{\begin{array}{ll}
\mathbf{A}_{i,2} = \delta^3_i\mathbf{A}_{1} + \delta^4_i\mathbf{A}_{2},\quad \cr
\mathbf{A}_{i,3}= \delta^3_i\mathbf{A}_{2} - \delta^4_i\mathbf{A}_{1},\cr
\mathbf{A}_{i,1}-u^2\mathbf{A}_{i,2} - u^3\mathbf{A}_{i,3} =  \delta^1_i\mathbf{A}_{1} + \delta^2_i\mathbf{A}_{2},  \cr
u^3\mathbf{A}_{i,2}-u^2\mathbf{A}_{i,3} - \mathbf{A}_{i,4} =  \delta^1_i\mathbf{A}_{2} - \delta^2_i\mathbf{A}_{1}.
\\\end{array}\ \right.\end{equation}
We integrate the system of equations \eqref{48}. As a result, we get the following solutions:
\begin{equation}\label{42} \left\{\begin{array}{ll}
\mathbf{A}_{1} = a_1 \cos(u^4+a) \exp(u^1), \cr
\mathbf{A}_{2} = a_1 \sin(u^4+a) \exp(u^1), \cr
\mathbf{A}_{3} = -a_3 + \alpha_1\exp(u^1)(u^2\cos(u^4+a) + u^3\sin(u^4+a)),\cr
\mathbf{A}_{4} = \alpha_1\exp(u^1)(u^2\sin(u^4+a)-u^3\cos(u^4+a)) +\alpha_4. \\\end{array}\ \right.
\end{equation}
Let us denote \quad $\alpha_1 = -a_1\sin a, \quad \alpha_1 = a_1\cos a, \quad  \alpha_3 = a_1. $ \quad Then the holonomic components of the vector potential \quad $A_i =\xi_i^\alpha \mathbf{A}_{\alpha}$,\quad and the matrixes \quad $||e^\alpha_i||, \quad ||e_\alpha^i||$ \quad have the form:
\begin{equation}\label{43}
A_{1} =\alpha_1, \quad A_{2}=(\alpha_2\cos u^4 + \alpha_3\sin u^4) \exp(u^1), \quad
\end{equation}
$$
A_{3}= (\alpha_2 \sin u^4 - \alpha_3\cos u^4) \exp(u^1), \quad A_{4} =\alpha_4,
$$
 \begin{equation}\label{44}
||e^\alpha_i||=\begin{pmatrix}1 & 0 & 0 & 0
\\0 &\cos u^4 \exp(u^1)&\sin u^4 \exp(u^1) & 0 \\ 0 & \sin u^4 \exp(u^1)& -\cos u^4 \exp(u^1) &0 \\ 0 & 0 & 0 & 1 \end{pmatrix},
\end{equation}
$$
||e_\alpha^i||=\begin{pmatrix}1 & 0 & 0 & 0
\\0 &\cos u^4 \exp(-u^1)&\sin u^4 \exp(-u^1) & 0 \\ 0 & \sin u^4 \exp(-u^1)& -\cos u^4 \exp(-u^1) &0 \\ 0 & 0 & 0 & 1 \end{pmatrix}.
$$

\subsection{Solvable $G_4$ containing an Abelian subgroups $G_3$}
The components of the Killing vector fields $\xi_\alpha^i$  and the components of the dual tetrad \quad $\xi^\alpha_i$ for groups containing a three-parameter Abelian subgroup can be represented as:
\begin{equation}\label{45}
||\xi_\alpha^i||=\begin{pmatrix}1 & 0 & 0 & 0
\\0 & 1 & 0 & 0 \\ 0 & 0 & 1 & 0 \\ C_p^1 u^p & C_p^2 u^p & C_p^3 u^p & 1 \end{pmatrix}, \quad
||\xi^\alpha_i||=\begin{pmatrix}1 & 0 & 0 & 0
\\0 & 1 & 0 & 0 \\ 0 & 0 & 1 & 0 \\ -C_p^1 u^p & -C_p^2 u^p & -C_p^3 u^p & 1 \end{pmatrix},
\end{equation}
where \quad $C_p^q=const.$ \quad For each of the corresponding groups given in A.Z. Petrov \cite{13} the $C_p^q$ matrices have the following forms (below \quad $\varepsilon =0, 1; \quad l, k =const$).

\quad

 1. Group $G_4(VI_1)$:
 \begin{equation}\label{46}
 ||C_a^b||=\begin{pmatrix}l & 0 & 0
\\0 & \varepsilon & 0\\ 0 & 0 & k\end{pmatrix}.
 \end{equation}

 2. Group $G_4(VI_2)$:
 \begin{equation}\label{47}
 ||C_a^b||=\begin{pmatrix}l & 0 & 0
\\0 & k & 1\\ 0 & -1 & k\end{pmatrix}.
 \end{equation}

 3. Group $G_4(VI_3)$:
 \begin{equation}\label{48}
 ||CC_a^b||=\begin{pmatrix}\varepsilon & 0 & 0
\\0 & k & 1\\ 0 & 0 & k\end{pmatrix}.
 \end{equation}

 4. Group $G_4(VI_4)_1$:
 \begin{equation}\label{49}
 ||C_a^b||=\begin{pmatrix}\varepsilon & 0 & 0
\\0 & k & 1\\ 1 & 0 & k\end{pmatrix},\quad k \ne \varepsilon.
 \end{equation}

 5. Group $G_4(VI_4)_2$:
 \begin{equation}\label{50}
 ||C_a^b||=\begin{pmatrix}k & 0 & 0
\\0 & k & 1\\ 1 & 0 & k\end{pmatrix}.
 \end{equation}
 The system of equations \eqref{12} for the groups \eqref{46} - \eqref{50} can be represented as:
\begin{equation}\label{51}
\mathbf{A}_{i,a} = \mathbf{A}_{4,4} = 0, \quad \Rightarrow \mathbf{A}_a = \mathbf{A}_a(u^4), \quad \mathbf{A}_4 = \alpha_4. \quad
\end{equation}
\begin{equation}\label{52}
\mathbf{A}_{a,4} = -C_a^b \mathbf{A}_b.
\end{equation}
It can be shown that the \eqref{59} equations are satisfied only at zero electromagnetic field. Indeed, the holonomic components of the vector potential have the form:
\begin{equation}\label{53}
A_a = \mathbf{A}_a, \quad A_4 = -C^a_b u^b \mathbf{A}_a.
\end{equation}
That is why:
$$
F_{ab} = 0, \quad F_{a4} =  \mathbf{A}_{4,a} - \mathbf{A}_{a,4} =C_a^b \mathbf{A}_b -C_a^b \mathbf{A}_b = 0.
$$

\subsection{Unsolvable groups}
It is convenient to start the integration of the equations \eqref{12} by considering an analogous problem for the insoluble group $G_3(VIII)$,  $G_3(IX)$. Then we add a fourth integral of motion commuting with the operators of the unsolvable groups $G_3$. This operator can take the following forms:

a) $X_4=p_4.$

b) $X_4 = p_1 + p_4$.

\subsubsection{Unsolvable group $G_4(VII)_1$}

Let us integrate the system of equations \eqref{12} for the group $G_3(VIII)$.
The components of the Killing vectors \quad $\xi_\alpha^i$ \quad and the components of the dual tetrad \quad $\xi^\alpha_i$  \quad as well as non-zero components of structure constants \quad $C^\alpha_{\beta\gamma}$\quad of the group can $G_3(VII)$ be represent in the form  $ (a, b = 1, 2, 3)$:
\begin{equation}\label{54}
||\xi^\alpha_a||=\begin{pmatrix}0 & 1  & 0
\\0 & u^2 & 1 \\ \exp(u^3)  & {u^2}^2 & 2u^2 \end{pmatrix},
\end{equation}
$$\quad||\xi_\alpha^a||=\begin{pmatrix}{u^2}^2\exp(-u^3)& -2{u^2}\exp(-u^3)  & \exp(-u^3)\\1 & 0 & 0\\ -{u^2} & 1  & 0 \\\end{pmatrix};$$
$$
C^1_{12} = 1, \quad C^3_{23} = 1, \quad C^2_{13}=2.
$$
Let's find the system of defining equations \eqref{12}:
\begin{equation}\label{55}
\left\{\begin{array}{ll}
\mathbf{A}_{1,1}=2(u^2\mathbf{A}_{1} - \mathbf{A}_{2})\exp(-u^3), \quad \mathbf{A}_{1,2}=0, \quad
\mathbf{A}_{1,3}= -\mathbf{A}_{1};\cr
\mathbf{A}_{2,1}=({u^2}^2\mathbf{A}_{1} - \mathbf{A}_{3})\exp(-u^3), \quad
\mathbf{A}_{2,2}= \mathbf{A}_{1}, \quad \mathbf{A}_{2,3}=-u^2\mathbf{A}_{1};\cr

\mathbf{A}_{3,1} = 2u^2({u^2}\mathbf{A}_{2} - \mathbf{A}_{3})\exp(-u^3), \quad

\mathbf{A}_{3,2}= 2 \mathbf{A}_{2}, \quad \mathbf{A}_{3,3}=\mathbf{A}_{3} -2u^2\mathbf{A}_{2}.

\\\end{array}\ \right.\end{equation}
We integrate the system of equations \eqref{60} and obtain the following solutions:
\begin{equation}\label{56} \left\{\begin{array}{ll}
\mathbf{A}_{1} = Q\exp(-u^3),\quad
\mathbf{A}_{2} = u^2Q\exp(-u^3)-(a_1 u^1 +a_2),\cr \mathbf{A}_{3} ={u^2}^2Q\exp(-u^3)-u^2(a_1 u^1 +a_2) + a_1\exp{u^3}, \cr \mathbf{A}_4 = a_4, \quad Q = a_1{u^1}^2 - 2a_2u^1 + a_3,
\quad  a_a=a_a(u^4) \\\end{array}\ \right.\end{equation}. This solution has been found in \cite{17}, \cite{45}. The holonomic components of the vector potential \quad $A_i =\xi_i^\alpha \mathbf{A}_{\alpha}$,\quad and the matrixes \quad $||e^\alpha_i||, \quad ||e_\alpha^i||$ \quad have the form:
$$
A_{1} =\alpha_1, \quad A_{2}=Q \exp(-u^3), \quad A_{3}= -a_1 {u^1} + a_2, \quad A_{4} =a_4,
$$
\begin{equation}\label{57}
||e^\alpha_a||=\begin{pmatrix}1 & 0 & 0
\\{u^1}^2\exp{(-u^3)} &-2{u^1}\exp{(-u^3)} &\exp{(-u^3)}\\ -{u^1} & 1& 0 \end{pmatrix},
\end{equation}
$$
||e_\alpha^a||=\begin{pmatrix}1 & 0 & 0
\\{u^1} & 0 & 1\\ {u^1}^2 & \exp{u^3} & 2u^1 \end{pmatrix}.
$$
The metric tensor of the space with the group $G_3(VIII)$ is expressed in terms of $e_\alpha^i$ as follows:
\begin{equation}\label{58}
g^{ij}=\delta_4^i\delta_4^j + e_\alpha^i e_\beta^i \eta^{\alpha\beta}(u^4).
\end{equation}
To obtain a solution for the group $G_4(VII)$, we need to add the operator $X_4$. Then $a_a=\alpha_a, \quad \eta_{\alpha\beta} =const$. Let us consider possible variants.

\quad

a) $X_4=p_4.$ \quad The solution has the form:
\begin{equation}\label{59}
\quad||e_\alpha^i||=\begin{pmatrix}1 & 0 & 0 & 0
\\{u^1} & 0 & 1 & 0\\ {u^1}^2 & \exp{u^3} & 2u^1 & 0
\\0& 0& 0& 1\end{pmatrix}.
\end{equation}
$$
||e^\alpha_i||=\begin{pmatrix}1 & 0 & 0 &0
\\{u^1}^2\exp{(-u^3)} &-2{u^1}\exp{(-u^3)} &\exp{(-u^3)} &0\\ -{u^1} & 1& 0 &0
\\0& 0& 0& 1
\end{pmatrix},
$$

b) $X_4 = p_1 + p_4.$
Let us denote:
$$
\tilde {u}^1=u^1-u^4.
$$
Then the solution can be represent in the form:
$$
A_{1} = \alpha_1, \quad A_{2}=(\alpha_1 (\tilde{u}^1)^2 - 2\alpha_2 \tilde{u}^1 +\alpha_3),$$
$$
\quad A_{3}= \alpha_1 \tilde{u}^1 -u^2(\alpha_1 (\tilde{u}^1)^2 - 2\alpha_2 \tilde{u}^1 +\alpha_3)-\alpha_2. \quad A_{4} = \alpha_4,
$$
\begin{equation}\label{60}
||e_\alpha^i||=\begin{pmatrix}1 & 0 & 0 & 0
\\\tilde{u}^1 & 0 & 1 & 0\\ (\tilde{u}^1)^2 & \exp{u^3} & 2\tilde{u}^1 & 0
\\1 & 0& 0& 1
 \end{pmatrix},
\end{equation}
$$
||e^\alpha_i||=\begin{pmatrix}1 & 0 & 0 &0
\\ \tilde{u}^2\exp{(-u^3)} &-2\tilde{u}^1\exp{(-u^3)} &\exp{(-u^3)} &0\\ -\tilde{u}^1 & 1& 0 &0
\\-1& 0& 0& 1
\end{pmatrix},
$$

\subsection{Unsolvable group $G_4(VIII)$}

Let us integrate the system of equations \eqref{12} for the group $G_3(IX)$.
The components of the Killing vectors \quad $\xi_\alpha^i$ \quad and the components of the dual tetrad \quad $\xi^\alpha_i$  \quad as well as non-zero components of structure constants \quad $C^\alpha_{\beta\gamma}$\quad of the group can $G_3(IX)$ be represent in the form:
\begin{equation}\label{61}
||\xi_\alpha^a||=\begin{pmatrix}0
& 1 & 0
\\\cos{u^2} & -\frac{\sin u^2\cos u^1}{\sin u^1}  & \frac{\sin u^2}{\sin u^1}
\\-\sin{u^2}& -\frac{\cos u^2\cos u^1}{\sin u^1} & \frac{\cos u^2}{\sin u^1} \end{pmatrix},
\end{equation}
$$||\xi^\alpha_a||=\begin{pmatrix}
0& \cos u^2  & -\sin u^2
\\1 & 0 & 0\\\cos u^1& \sin u^2\sin u^1 & \cos u^2\sin u^1
\end{pmatrix}, \quad
C^3_{12} = 1, \quad C^1_{23} = 1, \quad C^2_{31}=1.
$$
Let's find the system of equations \eqref{12}:
\begin{equation}\label{62}
\left\{\begin{array}{ll}
\mathbf{A}_{1,1}\cos{u^2} +\mathbf{A}_{1,3}\frac{\sin{u^2}}{\sin{u^1}} = -\mathbf{A}_{3}, \cr
\mathbf{A}_{1,1}\sin{u^2} -\mathbf{A}_{1,3}\frac{\cos{u^2}}{\sin{u^1}} = -\mathbf{A}_{2} , \quad \mathbf{A}_{1,2}=0;\cr
\mathbf{A}_{2,1}\cos{u^2} +\mathbf{A}_{2,3}\frac{\sin{u^2}}{\sin{u^1}} = \mathbf{A}_{3}\frac{\cos{u^1}\sin{u^2}}{\sin{u^1}}, \cr
\mathbf{A}_{2,1}\sin{u^2} -\mathbf{A}_{2,3}\frac{\cos{u^2}}{\sin{u^1}} = \mathbf{A}_{1} - \mathbf{A}_{3}\frac{\cos{u^1}\sin{u^2}}{\sin{u^1}},\quad \mathbf{A}_{2,2}=\mathbf{A}_{3}, \cr
\mathbf{A}_{3,1}\cos{u^2} +\mathbf{A}_{3,3}\frac{\sin{u^2}}{\sin{u^1}} =\mathbf{A}_{1} - \mathbf{A}_{2}\frac{\cos{u^1}\sin{u^2}}{\sin{u^1}}, \cr
\mathbf{A}_{3,1}\sin{u^2} -\mathbf{A}_{3,3}\frac{\cos{u^2}}{\sin{u^1}} = \mathbf{A}_{2}\frac{\cos{u^1}\cos{u^2}}{\sin{u^1}}, \quad \mathbf{A}_{3,2}=-\mathbf{A}_{2}.
\\\end{array}\ \right.\end{equation}
We integrate the system of equations \eqref{65}. As a result, we get the following solutions:
\begin{equation}\label{63} \left\{\begin{array}{ll}
\mathbf{A}_{1} =  \alpha_3\cos u^1 +\alpha\sin u^1\sin u , \quad  \cr
\mathbf{A}_{2} = -\mathbf{A}_{1,1}\sin u^2  + \mathbf{A}_{1,3}\frac{\cos u^2}{\sin u^1}, \cr
\mathbf{A}_{3} = -\mathbf{A}_{1,1}\sin u^2  - \mathbf{A}_{1,3}\frac{\cos u^2}{\sin u^1}, \cr
\mathbf{A}_{4} =a_4. \\\end{array}\ \right.
\end{equation}
The holonomic components of the vector potential \quad $A_i =\xi_i^\alpha \mathbf{A}_{\alpha}$,\quad and the matrixes \quad $||e^\alpha_a||, \quad ||e_\alpha^a||$ have the form (see \cite{17}, \cite{46}):
\begin{equation}\label{64}
A_{1} = a_1\cos{u}, \quad A_{2}=a_3\cos u^1 + a_1\sin u^1 \sin{u},
\end{equation}
$$
 A_{3}= a_3, \quad A_{4} =a_4 \quad (u=u^3 + a_2, \quad a_a=a_a(x^4));
$$
 \begin{equation}\label{65}
 e_a^\alpha=\begin{pmatrix} \cos{u^3} &-\sin{u^3}  & 0 \\ \sin{u^1}\sin{u^3} & \sin{u^1}\cos{u^3}  & \cos{u^1} \\ 0 & 0 & 1  \end{pmatrix},
\end{equation}
$$
||e_\alpha ^a||=\begin{pmatrix} \cos{u^3} &\frac{\sin{u^3}}{\sin{u^1}} & -\frac{\sin{u^3}\cos{u^1}}{\sin{u^1}} \\-\sin{u^3} & \frac{\cos{u^3}}{\sin{u^1}}  & -\frac{\cos{u^3} \cos{u^1}}{\sin{u^1}} \\0 & 0 & 1  \end{pmatrix}.
$$
The metric tensor of the space with the group $G_3(IX)$ is expressed in terms of $e_\alpha^i$ as follows:
\begin{equation}\label{66}
g^{ij}=\delta_4^i\delta_4^j + e_\alpha^i e_\beta^j \eta^{\alpha\beta}(u^4).
\end{equation}
To obtain a solution for the group $G_4(VIII)$, we need to add the operator $X_4$. Then \quad $a_a=\alpha_a, \quad \eta_{\alpha\beta} =const, \quad$ Let us consider possible variants.

\quad

a) $X_4=p_4.$  The solution is given by the formulas \eqref{62}, \eqref{64} under the conditions:
\begin{equation}\label{67}
 ||e_\alpha ^i||=\begin{pmatrix} \cos{u^3} &\frac{\sin{u^3}}{\sin{u^1}} & -\frac{\sin{u^3}\cos{u^1}}{\sin{u^1}} & 0 \\-\sin{u^3} & \frac{\cos{u^3}}{\sin{u^1}}  & -\frac{\cos{u^3} \cos{u^1}}{\sin{u^1}} & 0\\0 & 0 & 1 & 0 \\ 0 & 0 & 0 & 1 \end{pmatrix}.
\end{equation}

b) $X_4 = p_3 + p_4.$
Let us denote:
$$
\tilde {u}^3=u^3-u^4.
$$
Then the solution can be represent in the form:
$$
A_{1} = \alpha_1\cos{\tilde{u}^3}, \quad A_{2}=\alpha_3\cos u^1 + \alpha_1\sin u^1 \sin{\tilde{u}^3} , \quad A_{3}= \alpha_3, \quad A_{4} =\alpha_4;
$$
\begin{equation}\label{68}
 ||e_\alpha ^i||=\begin{pmatrix} \cos{\tilde{u}^3} &\frac{\sin{\tilde{u}^3}}{\sin{u^1}} & -\frac{\sin{\tilde{u}^3}\cos{u^1}}{\sin{u^1}} & 0 \\-\sin{\tilde{u}^3} & \frac{\cos{\tilde{u}^3}}{\sin{u^1}}  & -\frac{\cos{\tilde{u}^3} \cos{u^1}}{\sin{u^1}} & 0\\0 & 0 & 1 & 0  \\ 0 & 0 & 1 & 1 \end{pmatrix},
\end{equation}
$$
||e_i^\alpha||=\begin{pmatrix} \cos{\tilde{u}^3} &-\sin{\tilde{u}^3}  & 0 & 0\\ \sin{u^1}\sin{\tilde{u}^3} & \sin{u^1}\cos{\tilde{u}^3} & \cos{u^1} & 0 \\ 0 & 0 & 1 & 0 \\0 & 0  & -1 & 1 \end{pmatrix}.
$$

Note that the tetrad \eqref{60}, \eqref{68} define spaces $V_4$ on which unsolvable isometry groups $G_4$ acts simply transitively. Therefore, Petrov's classification should be completed with the following spaces.

1. The space $V_4$ on which the group $G_4(VII)$ acts simply transitively. The group operators are of the form:
\begin{equation}\label{69}
X_1 = \exp(-u^3)(p_1 - {u^2}^2p_2 - 2u^2), \quad X_2 = p_3 , \quad X_3 =\exp(u^3)p_2, \quad X_4=p_4.
\end{equation}
In Petrov's formula (26.34), it is sufficient to provide for the possibility of turning the parameter $\varepsilon $ to zero.

2. The space $V_4$ on which the group $G_4(VIII)$ acts simply transitively. The group operators are of the form:
\begin{equation}\label{70}
X_1 = p_1\cos{\tilde{u}^3} + \frac{\sin\tilde{u}^3}{\sin u^1}(p_2 - p_3\cos u^1), \quad X_2 = \partial_3 X_1,\quad
\end{equation}
$$
X_4=p_4 - p_3, \quad \tilde{u}^3 =u^3 + u^4.
$$
In Petrov's formula (26.35), it is sufficient to choose the fourth operator in the form:
$$ X_4 =p_4 - \varepsilon p_3,
$$
where $\varepsilon = 0, 1$.

\section{Conclusion}

The results obtained in this article complete the decision of the classification problem for space-time metrics and potentials of admissible electromagnetic fields, in which the Hamilton-Jacobi equation for a charged test particle admits complete sets of integrals of motion. For the case of non-commutative complete sets this classification problem has been actually solved for the Klein-Gordon-Fock equation as well.

The next problem will the classification of exact solutions of the Einstein-Maxwell vacuum equations in spaces with a non-commuting complete sets and admissible electromagnetic fields using the obtained results. The first part of this  problem - the classification of exact solutions of Maxwell's vacuum equations has already been solved (see\cite{44}-\cite{47}).

\quad

FUNDING: The work is supported by Russian Science Foundation, project number N 23-21-00275.

\quad

\end{document}